# A new peridynamic formulation with shear deformation for elastic solid

Huilong Ren[a], Xiaoying Zhuang[b,c,*], Timon Rabczuk[a,d,e,*]

[a]*Institute of Structural Mechanics, Bauhaus-Universitt Weimar, 99423 Weimar, Germany*
[b]*State Key Laboratory of Disaster Reduction in Civil Engineering, College of Civil Engineering,Tongji University, Shanghai 200092, China*
[c]*Institute of Conitnuum Mechanics, Leibniz University Hannover, Hannover, Germany*
[d]*Division of Computational Mechanics, Ton Duc Thang University, Ho Chi Minh City, Viet Nam*
[e]*Faculty of Civil Engineering, Ton Duc Thang University, Ho Chi Minh City, Viet Nam*

**Abstract**

We propose a new peridynamic formulation with shear deformation for linear elastic solid. The key idea lies in subtracting the rigid body rotation part from the total deformation. Based on the strain energy equivalence between classic local model and non-local model, the bond force vector is derived. A new damage rule of maximal deviatoric bond strain for elastic brittle fracture is proposed in order to account for both the tensile damage and shear damage. 2D and 3D numerical examples are tested to verify the accuracy of the current peridynamics. The new damage rule is applied to simulate the propagation of Mode I, II and III cracks.

## 1. Introduction

The prediction of crack is always an interesting topic in the field of computational solid mechanics. Various numerical methods have been proposed, such as the Element-free Galerkin methods (EFG) [1], the extended finite element method (XFEM)[2, 3], the reproducing kernel particle method (RKPM)[4], the numerical manifold method (NMM)[5, 6], cracking particle methods (CPM)[7], and many other meshless methods [8, 9, 10]. Peridynamics [11] has recently attracted great attention due to its flexibility in modeling complex fracture patterns. Peridynamic theory reformulates the problems in solid mechanics in terms of integral form rather than the partial differential form, thus avoids the singularities arose at the crack tips and discontinuity in differential across the cracks.

*Institute of Conitnuum Mechanics, Leibniz University Hannover, Hannover, Germany; Institute of Structural Mechanics, Bauhaus-Universit at Weimar, Germany. E-mail: zhuang@ikm.uni-hannover.de; timon.rabczuk@uni-weimar.de; timon.rabczuk@tdt.edu.vn

The derivation of peridynamics is based on the equivalence of local strain energy density and the nonlocal strain energy density [12, 13]. There are general three types of peridynamics, bond-based (BB) peridynamics, ordinary state based (OSB) peridynamics and non-ordinary state based (NOSB) peridynamics. The NOSB peridynamics incorporates various advanced material models [14, 15, 16, 17] as it is based on the deformation gradient in continuum mechanics. Foster *et al* [14] incorporated the viscoplasticity in NOSB peridynamics. Amani *et al* [17] implemented the Johnson-Cook (JC) and damage model based on NOSB peridynamics. Apart from the applications for solid, peridynamics has also been extended to other fields, heat diffusion [18, 19, 20], plates and shells [21, 22, 23], composite materials [24] and the nonlocal vector calculus [25].

The most distinguished advantage of peridynamics over other numerical methods is that the fracture is a natural result of simulation, whereas the techniques such as the smoothing of normals of crack surfaces used in the extended finite element method (XFEM) [26], meshless methods [27] or other partition of unity methods (PUM) [28] are not needed. The drawback of NOSB peridynamics is the relatively complicated implementation and high computational cost. The capability of NOSB peridynamics is attributed to the arbitrary direction of bond force, which is in contrast to the BB peridynamics or OSB peridynamics only allowing extensional bond force. In fact, the component of bond force out of the bond direction in NOSB peridynamics represents the shear force due to shear deformation. It can be seen that one problem in BB peridynamics or OSB peridynamics is that the axial bond force cannot account for the shear deformation but only extensional deformation. Such limitation is physically unnatural. More specifically, for deformed solid, there are both shear and axial deformations, the bond should bear some energy due to shear deformation, not just the energy due to extensional deformation. Combining the arbitrary direction of bond force in NOSB peridynamics and the relatively low computational cost in OSB peridynamics, we can formulate a new peridynamic formulation with shear deformation for elastic solid.

In this paper, we present a new peridynamic formulation with shear deformation for linear elastic solid. The new approach is based on subtracting the part of rigid body rotation from the total deformation since the pure rotation does not contribute to the internal force. The content of the paper is outlined as follows. §2 begins with the equations of motion of dual-horizon peridynamics. In §3 the bond force vector for the present peridynamic formulation is derived. §4 recovers the Cauchy stress tensor based on the bond force vector proposed in §3. §5 discusses the shortcomings of the conventional damage rule of maximal extension stretch and a new damage rule is proposed based on deviatoric bond strain which takes into account both the tensile damage and shear damage in brittle fracture. In §6, four numerical examples are presented to validate the present formulation.

## 2. Dual-horizon peridynamics

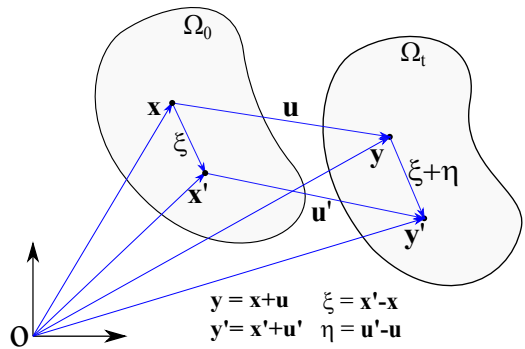

Figure 1: The configuration for deformed body.

Consider a solid body in its initial configuratoin before deformatoin and current configuration after deformation as shown in Fig. 1. Let $\mathbf{x}$ and $\mathbf{x}'$ be the material coordinates of any two points in the initial configuration $\mathbf{\Omega}_0$; and $\mathbf{y} := \mathbf{y}(\mathbf{x}, t)$ and $\mathbf{y}' := \mathbf{y}(\mathbf{x}', t)$ are the spatial coordinates of the two points in the current configuration $\mathbf{\Omega}_t$; $\boldsymbol{\xi} := \mathbf{x}' - \mathbf{x}$ denotes the initial bond vector, i.e. the relative distance vector between $\mathbf{x}$ and $\mathbf{x}'$, and $\mathbf{u} := \mathbf{u}(\mathbf{x}, t)$ and $\mathbf{u}' := \mathbf{u}(\mathbf{x}', t)$ are the displacement vectors for $\mathbf{x}$ and $\mathbf{x}'$, respectively. Hence, $\boldsymbol{\eta} := \mathbf{u}' - \mathbf{u}$ is the relative displacement vector for bond $\boldsymbol{\xi}$. $\mathbf{y}\langle\boldsymbol{\xi}\rangle := \mathbf{y}(\mathbf{x}',t) - \mathbf{y}(\mathbf{x},t) = \boldsymbol{\xi} + \boldsymbol{\eta}$ is the current bond vector for bond $\boldsymbol{\xi}$. The horizon of point $\mathbf{x}$ in traditional peridynamics represents a domain centered at $\mathbf{x}$, in which the distance from any point $\mathbf{x}'$ to $\mathbf{x}$ is smaller than $\delta$, the radius of horizon. In the traditional peridynamics, the horizon sizes for all the particles must be the same, otherwise spurious wave reflections will arise [29]. In order to remove the constraints of using constant horizon, much effort were devoted by [30, 31, 32, 29], among which the simplest method is probably the dual-horizon peridynamics [29]. The key idea of this method is the introduction of dual-horizon, the dual term of the horizons centering at each particle. Within the framework of dual-horizon peridynamics, the traditional peridynamics with constant horizon can be derived as a special case of dual-horizon peridynamics. In the following sections, the derivation will be based on dual-horizon peridynamics.

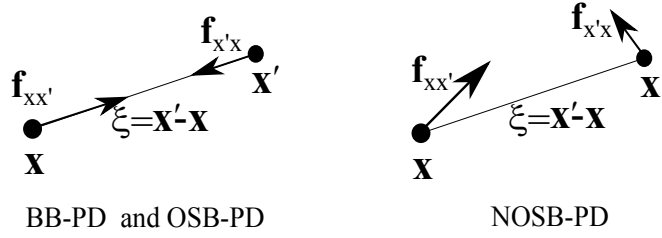

Figure 2: The bond force vectors in dual-horizon peridynamics.

2.1. Equations of motions in dual-horizon peridynamics

In dual-horizon peridynamic formulation, the *horizon* $H_\mathbf{x}$ is defined as the domain where any particle falling inside will receive force exerted by $\mathbf{x}$, therefore, $\mathbf{x}$ will undertake all the reaction forces from the particles inside $H_\mathbf{x}$. In this sense, the horizon in dual-horizon peridynamics can be viewed as a reaction force horizon. The *dual-horizon* in dual-horizon peridynamics is defined as a union of points whose horizons include $\mathbf{x}$, denoted by

$$H'_\mathbf{x} = \{\mathbf{x}'|\mathbf{x} \in H_{\mathbf{x}'}\} \tag{1}$$

The particle will receive the direct force from the particles in its dual-horizon. The dual-horizon can be viewed as direct force horizon. Let $\mathbf{f}_{\mathbf{x}'\mathbf{x}} := \mathbf{f}_{\mathbf{x}'\mathbf{x}}(\boldsymbol{\eta},\boldsymbol{\xi})$ denote the force vector density acting on particle $\mathbf{x}'$ due to particle $\mathbf{x}$. The first subscript $\mathbf{x}'$ of $\mathbf{f}_{\mathbf{x}'\mathbf{x}}$ indicates $\mathbf{x}'$ being the object of force and the second subscript $\mathbf{x}$ indicates the source of force which is from $\mathbf{x}$.

The internal forces exerted at each particle include two parts, the reaction forces from the horizon and the direct forces from the dual-horizon. The other forces applied to a particle include the body force and the inertia. Let $\Delta V_{\mathbf{x}}$ denote the volume associated with $\mathbf{x}$. The body force of particle $\mathbf{x}$ can be expressed as $\rho_{\mathbf{x}}\mathbf{b}\Delta V_{\mathbf{x}}$, where $\mathbf{b}$ is the body force density, e.g. the gravity. The inertia is denoted by $\rho_{\mathbf{x}}\ddot{\mathbf{u}}(\mathbf{x},t)\Delta V_{\mathbf{x}}$, where $\rho_{\mathbf{x}}$ is the averaged density of the volume associated with $\mathbf{x}$.

The direct force is

$$\mathbf{f}_{\mathbf{x}\mathbf{x}'}(-\boldsymbol{\eta},-\boldsymbol{\xi}) \cdot \Delta V_{\mathbf{x}} \cdot \Delta V_{\mathbf{x}'}, \forall \mathbf{x}' \in H'_{\mathbf{x}} \tag{2}$$

The reaction force is

$$-\mathbf{f}_{\mathbf{x}'\mathbf{x}}(\boldsymbol{\eta},\boldsymbol{\xi}) \cdot \Delta V_{\mathbf{x}'} \cdot \Delta V_{\mathbf{x}}, \forall \mathbf{x}' \in H_{\mathbf{x}} , \tag{3}$$

By summing over all forces on the volume associated with particle $\mathbf{x}$, including inertia, body force, direct force in Eq. (2) and reaction force in Eq. (3), the equation of motion for dual-horizon peridynamics is obtained,

$$\rho_{\mathbf{x}}\ddot{\mathbf{u}}(\mathbf{x},t)\Delta V_{\mathbf{x}} = \sum_{\mathbf{x}' \in H'_{\mathbf{x}}} \mathbf{f}_{\mathbf{x}\mathbf{x}'}(-\boldsymbol{\eta},-\boldsymbol{\xi})\Delta V_{\mathbf{x}'}\Delta V_{\mathbf{x}} + \sum_{\mathbf{x}' \in H_{\mathbf{x}}} (-\mathbf{f}_{\mathbf{x}'\mathbf{x}}(\boldsymbol{\eta},\boldsymbol{\xi})\Delta V_{\mathbf{x}'}\Delta V_{\mathbf{x}}) + \rho_{\mathbf{x}}\mathbf{b}\Delta V_{\mathbf{x}} . \tag{4}$$

As the volume $\Delta V_{\mathbf{x}}$ associated with particle $\mathbf{x}$ is independent of the summation, we can eliminate $\Delta V_{\mathbf{x}}$ in Eq. (4), and arrive at the governing equation based on $\mathbf{x}$:

$$\rho_{\mathbf{x}}\ddot{\mathbf{u}}(\mathbf{x},t) = \sum_{\mathbf{x}' \in H'_{\mathbf{x}}} \mathbf{f}_{\mathbf{x}\mathbf{x}'}(-\boldsymbol{\eta},-\boldsymbol{\xi})\Delta V_{\mathbf{x}'} - \sum_{\mathbf{x}' \in H_{\mathbf{x}}} \mathbf{f}_{\mathbf{x}'\mathbf{x}}(\boldsymbol{\eta},\boldsymbol{\xi})\Delta V_{\mathbf{x}'} + \rho_{\mathbf{x}}\mathbf{b} . \tag{5}$$

When the discretisation is sufficiently fine, the summation is approximating the integration of the force over the dual-horizon and horizon,

$$\lim_{\Delta V_{\mathbf{x}'} \to 0} \sum_{\mathbf{x}' \in H'_{\mathbf{x}}} \mathbf{f}_{\mathbf{x}\mathbf{x}'}(-\boldsymbol{\eta},-\boldsymbol{\xi})\Delta V_{\mathbf{x}'} = \int_{\mathbf{x}' \in H'_{\mathbf{x}}} \mathbf{f}_{\mathbf{x}\mathbf{x}'}(-\boldsymbol{\eta},-\boldsymbol{\xi})\,\mathrm{d}V_{\mathbf{x}'}$$

$$\lim_{\Delta V_{\mathbf{x}'} \to 0} \sum_{\mathbf{x}' \in H_{\mathbf{x}}} \mathbf{f}_{\mathbf{x}'\mathbf{x}}(\boldsymbol{\eta},\boldsymbol{\xi})\Delta V_{\mathbf{x}'} = \int_{\mathbf{x}' \in H_{\mathbf{x}}} \mathbf{f}_{\mathbf{x}'\mathbf{x}}(\boldsymbol{\eta},\boldsymbol{\xi})\,\mathrm{d}V_{\mathbf{x}'} .$$

Thus the integration form of the equation of motion in dual-horizon peridynamics is given as

$$\rho_{\mathbf{x}}\ddot{\mathbf{u}}(\mathbf{x},t) = \int_{\mathbf{x}'\in H'_{\mathbf{x}}} \mathbf{f}_{\mathbf{x}\mathbf{x}'}(-\boldsymbol{\eta},-\boldsymbol{\xi})\,\mathrm{d}V_{\mathbf{x}'} - \int_{\mathbf{x}'\in H_{\mathbf{x}}} \mathbf{f}_{\mathbf{x}'\mathbf{x}}(\boldsymbol{\eta},\boldsymbol{\xi})\,\mathrm{d}V_{\mathbf{x}'} + \rho_{\mathbf{x}}\mathbf{b}\ . \qquad (6)$$

## 3. Peridynamics with shear bond force

### 3.1. Geometric description

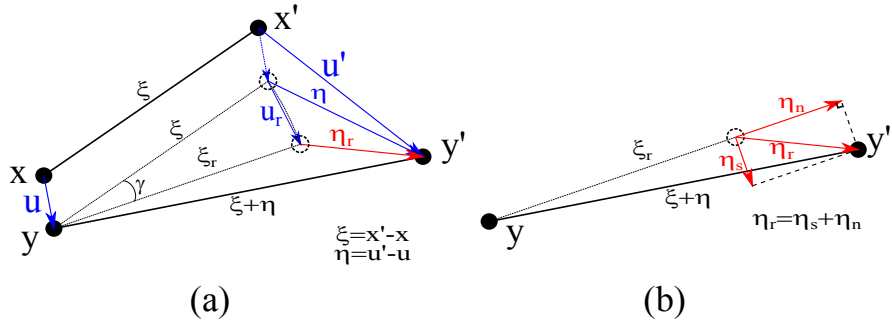

Figure 3: (a) The bond configuration; $\mathbf{u}$ is the rigid body translation, $\boldsymbol{\xi}_r$ is the bond after rigid body rotation, $\gamma$ denotes the angle of rotation, $\boldsymbol{\eta}_r$ is the effective displacement due to deformation; $\boldsymbol{\xi}_r + \boldsymbol{\eta}_r = \boldsymbol{\xi} + \boldsymbol{\eta}$. (b) The decomposition of the effective displacement, $\boldsymbol{\eta}_r = \boldsymbol{\eta}_s + \boldsymbol{\eta}_n$, $\boldsymbol{\eta}_s$ is the shear displacement perpendicular to $\boldsymbol{\xi}_r$, $\boldsymbol{\eta}_t$ is the displacement along $\boldsymbol{\xi}_r$.

Two particles $\mathbf{x}, \mathbf{x}'$ form a bond $\boldsymbol{\xi}$ in the initial configuration at $t = 0$, at time $t = T$, $\boldsymbol{\xi}$ becomes $\boldsymbol{\xi} + \boldsymbol{\eta}$ in the current configuration, where relative displacement is $\boldsymbol{\eta}$.
As shown in Fig.3, the change of bond $\boldsymbol{\xi}$ comprises three parts:
(1) the translation $\mathbf{u}$ associated with $\mathbf{x}$;
(2) the displacement $\mathbf{u}_r$ due to $\mathbf{x}'$ rigid rotation with respect to $\mathbf{x}$;
(3) the displacement $\boldsymbol{\eta}_r$ due to deformation, $\boldsymbol{\eta}_r = \boldsymbol{\xi} + \boldsymbol{\eta} - \boldsymbol{\xi}_r$.
In fact, the Taylor expansion of displacement field $\mathbf{u}'$ over $\mathbf{x}$ is

$$\mathbf{u}' \approx \mathbf{u} + \nabla\mathbf{u}\cdot\boldsymbol{\xi} = \mathbf{u} + \frac{1}{2}(\nabla\mathbf{u} - \mathbf{u}\nabla)\cdot\boldsymbol{\xi} + \frac{1}{2}(\nabla\mathbf{u} + \mathbf{u}\nabla)\cdot\boldsymbol{\xi}, \qquad (7)$$

where $\mathbf{u}\nabla = (\nabla\mathbf{u})^T$. The three terms in the R.H.S of Eq.(7) are the rigid body translation, rigid body rotation, and the effective displacement between $\mathbf{x}$ and $\mathbf{x}'$, respectively.

The infinitesimal rotation tensor is defined as

$$A = \frac{1}{2}(\nabla \mathbf{u} - \mathbf{u}\nabla) \tag{8}$$

For infinitesimal deformation, the rotational displacement can be approximated by

$$\mathbf{u}_r = \frac{1}{2}(\nabla \mathbf{u} - \mathbf{u}\nabla) \cdot \boldsymbol{\xi} \tag{9}$$

However, when the rigid rotation is moderate or large, the above expression is no longer valid. The rotation matrix for each particle should be calculated.

As the rigid rotation is associated with the collective deformation of all particles in one horizon, it is necessary to calculate the deformation gradient first. The deformation gradient is calculated by

$$\mathbf{F}_\mathbf{x} = \int_{H_\mathbf{x}} \omega(\boldsymbol{\xi})(\boldsymbol{\xi} + \boldsymbol{\eta}) \otimes \boldsymbol{\xi} \mathrm{d}V_{\mathbf{x}'} \cdot \left( \int_{H_\mathbf{x}} \omega(\boldsymbol{\xi}) \boldsymbol{\xi} \otimes \boldsymbol{\xi} \mathrm{d}V_{\mathbf{x}'} \right)^{-1} \tag{10}$$

where $\omega(\boldsymbol{\xi})$ is the influence function. The next step is to find the rigid rotation in $\mathbf{F}_\mathbf{x}$. Based on linear algebra, any square real matrix $\mathbf{F}_\mathbf{x}$ can be written with polar decomposition

$$\mathbf{F}_\mathbf{x} = \mathbf{R}_\mathbf{x} \mathbf{P}_\mathbf{x} \tag{11}$$

where $\mathbf{R}_\mathbf{x}$ is a rotation matrix and $\mathbf{P}_\mathbf{x}$ is a positive-semidefinite Hermitian matrix. Based on the singular value decomposition of $\mathbf{F}_\mathbf{x}$, $\mathbf{F}_\mathbf{x} = W\Sigma V^T$, one has

$$\mathbf{P}_\mathbf{x} = V\Sigma V^T, \quad \mathbf{R}_\mathbf{x} = WV^T \tag{12}$$

where superscript $T$ denotes transpose of the matrix. So the rigid rotation matrix $\mathbf{R}_\mathbf{x}$ of $\mathbf{F}_\mathbf{x}$ can be obtained by singular value decomposition. The bond $\boldsymbol{\xi}$ after rigid rotation $\mathbf{R}_\mathbf{x}$ is given by

$$\boldsymbol{\xi}_r = \mathbf{R}_\mathbf{x} \cdot \boldsymbol{\xi} \tag{13}$$

The effective displacement is $\boldsymbol{\eta}_r = \boldsymbol{\xi} + \boldsymbol{\eta} - \boldsymbol{\xi}_r$. The displacement $\boldsymbol{\eta}_r$ is decomposed into $\boldsymbol{\eta}_n$ and $\boldsymbol{\eta}_s$, which are parallel and perpendicular to $\boldsymbol{\xi}_r$, respectively. In fact, in the case of isotropic material, $\boldsymbol{\eta}_n$ corresponds to the axial deformation and $\boldsymbol{\eta}_s$ to the shear deformation of the bond. The weighted volume for particle $\mathbf{x}$ is defined as

$$m_{\mathbf{x}} = \int_{H_{\mathbf{x}}} \omega(\boldsymbol{\xi})\boldsymbol{\xi}_r \cdot \boldsymbol{\xi}_r \, dV = \int_{H_{\mathbf{x}}} \omega(\boldsymbol{\xi})\boldsymbol{\xi} \cdot \boldsymbol{\xi} dV \qquad (14)$$

The volume strain is defined as

$$\theta_{\mathbf{x}} = \frac{3}{m_{\mathbf{x}}} \int_{H_{\mathbf{x}}} \omega(\boldsymbol{\xi})\, \boldsymbol{\eta}_r \cdot \boldsymbol{\xi}_r dV_{\mathbf{x}'} \qquad (15)$$

The Frechet derivative [13] of volume strain with respect to $\boldsymbol{\eta}_r$ is

$$\nabla \theta_{\mathbf{x}}(\boldsymbol{\eta}_r) = \frac{3}{m_{\mathbf{x}}} \omega(\boldsymbol{\xi})\, \boldsymbol{\xi}_r \qquad (16)$$

Note that the volume dilation $\theta_{\mathbf{x}}$ only depends on $\boldsymbol{\eta}^i$

The isotropic part of the displacement is

$$\boldsymbol{\eta}^i = \frac{\theta_{\mathbf{x}}}{3} \boldsymbol{\xi}_r \qquad (17)$$

The deviatoric part of the displacement is

$$\boldsymbol{\eta}^d = \boldsymbol{\eta}_r - \boldsymbol{\eta}^i \qquad (18)$$

The corresponding deviatoric bond strain $\boldsymbol{\varepsilon}^d$ is

$$\boldsymbol{\varepsilon}^d = \frac{\boldsymbol{\eta}^d}{\xi} = \frac{\boldsymbol{\eta}_r - \boldsymbol{\eta}^i}{\xi} \qquad (19)$$

3.2. Energy equivalence

In 3D case, the elastic potential energy for isotropic elastic material is

$$v_E = \frac{1}{2} K \theta^2 + \mu \bar{\varepsilon}_{ij}^2 \qquad (20)$$

The averaged stress $\sigma_{ii}$ and the deviatoric stress $\bar{\sigma}_{ij}$ are obtained by

$$\sigma_{ii} = \frac{\partial v_E}{\partial \theta} = K\theta, \; \bar{\sigma}_{ij} = \frac{\partial v_E}{\partial \bar{\varepsilon}_{ij}} = 2\mu \bar{\varepsilon}_{ij} \qquad (21)$$

where $\bar{\varepsilon}_{ij} = \varepsilon_{ij} - \delta_{ij}\frac{\theta}{3}$, $\theta = \varepsilon_{kk}$, $\bar{\sigma}_{ij} = \sigma_{ij} - \delta_{ij}\frac{\sigma_{kk}}{3}$, $K$ is the bulk modulus, $\mu$ is the shear modulus. For isotropic material, the normal strain causes only normal stress, while shear strain causes only shear stress; in other words, the shear component and normal component are uncoupled. If the coupling effect between shear part and normal part is considered, the constitution of anisotropic material can be derived. Here, we focus on the isotropic materials. Suppose an isotropic elastic material model is given by

$$W(\theta, \boldsymbol{\eta}^d) = \frac{K\theta^2}{2} + \frac{\alpha}{2} \int_H \omega(\boldsymbol{\xi}) \boldsymbol{\eta}^d \cdot \boldsymbol{\eta}^d \mathrm{d}V \tag{22}$$

where $\alpha$ is positive constant to be determined, and $\omega(\boldsymbol{\xi})$ is an influence function. The influence function plays a great role in the distribution of strain energy for all bonds. When $\omega(\boldsymbol{\xi}) = 1$, $\omega(\boldsymbol{\xi})\boldsymbol{\eta}^d \cdot \boldsymbol{\eta}^d \propto \xi^2$, which means the particle close to the boundary of the horizon has the most significant influence on the integration while the influence of the particle near the center is almost negligible. When $\omega(\boldsymbol{\xi}) = \xi^{-2}$, we can see $\omega(\boldsymbol{\xi})\boldsymbol{\eta}^d \cdot \boldsymbol{\eta}^d \propto 1$, which means each particle's influence is independent of $\xi$. $\omega(\boldsymbol{\xi}) = \xi^{-2}$ enables any particle falling inside the horizon the same order of magnitude of bond potential energy, which guarantees that the elastic potential energy is released gradually when breaking bond. If $\omega(\boldsymbol{\xi}) = 1$, the bond close to horizon's boundary has much greater potential energy than that of bond near the horizon's center, in that case, the elastic potential energy clusters on horizon's boundary. Therefore, it is recommended that $\omega(\boldsymbol{\xi}) = \xi^{-2}$.

The isotropic part of the bond force is

$$\begin{aligned}
\mathbf{f}^i &= \frac{\partial W}{\partial \boldsymbol{\eta}^i} \\
&= \frac{\partial W}{\partial \theta} \frac{\partial \theta}{\partial \boldsymbol{\eta}^i} \\
&= K\theta_{\mathbf{x}} \frac{3}{m_{\mathbf{x}}} \omega(\boldsymbol{\xi}) \boldsymbol{\xi}_r \\
&= \frac{3K\theta_{\mathbf{x}}}{m_{\mathbf{x}}} \omega(\boldsymbol{\xi}) \boldsymbol{\xi}_r
\end{aligned} \tag{23}$$

The deviatoric part of the bond force is the Frechet derivative of $W$ on $\boldsymbol{\eta}^d$,

$$\mathbf{f}^d = \frac{\partial W}{\partial \boldsymbol{\eta}^d} = \alpha \omega(\boldsymbol{\xi}) \boldsymbol{\eta}^d \tag{24}$$

In order to determine the coefficient $\alpha$, consider a deformation state without bulk deformation $\varepsilon_{kk} = 0$, the deviatoric bond displacement is $\boldsymbol{\eta}_d = \boldsymbol{\eta}_r = \boldsymbol{\varepsilon} \cdot \boldsymbol{\xi}_r$, and the strain energy due to deviatoric deformation is

$$W = \frac{\alpha}{2}\int_H \omega(\boldsymbol{\xi})\boldsymbol{\eta}_d \cdot \boldsymbol{\eta}_d \mathrm{d}V_{\mathbf{x}'}$$
$$= \frac{\alpha}{2}\int_H \omega(\boldsymbol{\xi})(\boldsymbol{\varepsilon}\cdot\boldsymbol{\xi}_r)\cdot(\boldsymbol{\varepsilon}\cdot\boldsymbol{\xi}_r)\mathrm{d}V_{\mathbf{x}'}$$
$$= \frac{\alpha}{6}\varepsilon_{ij}^2\int_H \omega(\boldsymbol{\xi})\boldsymbol{\xi}_r\cdot\boldsymbol{\xi}_r \mathrm{d}V_{\mathbf{x}'}$$
$$= \frac{\alpha m}{6}\varepsilon_{ij}^2 \qquad (25)$$

On the other hand, $W = \mu\varepsilon_{ij}^2$, then

$$\alpha = \frac{6\mu}{m} \qquad (26)$$

where $m$ is the weighted volume of the horizon.

Therefore, the bond force for the peridynamics with shear deformation is

$$\mathbf{f} = \frac{3\omega(\boldsymbol{\xi})}{m_{\mathbf{x}}}(K\theta_{\mathbf{x}}\boldsymbol{\xi}_r + 2\mu\boldsymbol{\eta}^d) \qquad (27)$$

where $K$ is the bulk modulus, $\mu$ the shear modulus. It is evident that the Eq.(27) is similar to the classical elastic mechanics, $\boldsymbol{\sigma} = K\theta + 2\mu\boldsymbol{\varepsilon}^d$. It can be seen that the bond direction is not required to be parallel with the current bond direction, in that sense, the current peridynamics can be regarded as one type of non-ordinary state based peridynamics. Following the same routines, the bond force of current peridynamics with shear deformation in 2D is

$$\mathbf{f} = \frac{2\omega(\boldsymbol{\xi})}{m_{\mathbf{x}}}(K\theta_{\mathbf{x}}\boldsymbol{\xi}_r + 2\mu\boldsymbol{\eta}^d). \qquad (28)$$

$\mathbf{f}$ is the bond force acted on $\mathbf{x}$ due to the relative deformation between $\mathbf{x}$ and $\mathbf{x}'$. The magnitude of $\mathbf{f}$ depends on the state of $\mathbf{x}$. The force direction of $\mathbf{f}$ is the same as the deformation direction, elongation results in tensile bond force and compression gives rise to compressive bond force. Hence, $\mathbf{f}$ is the bond force acted on $\mathbf{x}$ due to $\mathbf{x}$; within the framework of dual-horizon peridynamics, $\mathbf{f}$ corresponds to $-\mathbf{f}_{\mathbf{x}'\mathbf{x}}$. So the bond force $\mathbf{f}_{\mathbf{x}'\mathbf{x}}$ and

$\mathbf{f_{xx'}}$ in Eq.(5) are given by

$$\mathbf{f_{x'x}} = -\frac{3\omega(\boldsymbol{\xi})}{m_\mathbf{x}}(K\theta_\mathbf{x}\boldsymbol{\xi}_r + 2\mu\boldsymbol{\eta}^d), \forall \mathbf{x}' \in H_\mathbf{x} \qquad (29)$$

$$\mathbf{f_{xx'}} = -\frac{3\omega(\boldsymbol{\xi})}{m_{\mathbf{x}'}}(K\theta_{\mathbf{x}'}\boldsymbol{\xi}'_r + 2\mu\boldsymbol{\eta}'^d), \forall \mathbf{x}' \in H'_\mathbf{x}. \qquad (30)$$

Note that $\boldsymbol{\xi}'_r$ and $\boldsymbol{\eta}'^d$ are calculated based on the volume strain and rigid rotation of $\mathbf{x}'$, and $\boldsymbol{\xi}'_r$ is nearly in opposite direction of $\boldsymbol{\xi}_r$. When calculating bond forces in $\mathbf{x}$, it is not necessary to calculate $\mathbf{f_{xx'}}$ directly; in fact, $\mathbf{f_{xx'}}$ can determined when calculating the bond force in $\mathbf{x}'$. It can be seen that the dual-horizon in dual-horizon peridynamics is not stored explicitly but inferred from horizons. Compared with the OSB peridynamics in Eq.(31), the feature of current peridynamic formulation is that the shear deformation causes shear bond force. It can be seen that the current peridynamics is vector-valued, while the OSB peridynamics is scalar-valued. In traditional BB peridynamics and OSB peridynamics [13], the shear deformation is neglected. The energy equivalence with classic theory forces the bond stretch deformation in OSB peridynamics higher energy density, i.e. the coefficient of deviatoric part is 15 in OSB peridynamics and 6 in current peridynamics.

$$\underline{t} = \frac{3K\theta_\mathbf{x}}{m_\mathbf{x}}\omega(\boldsymbol{\xi})\xi + \frac{15\mu}{m_\mathbf{x}}\omega\langle\boldsymbol{\xi}\rangle\underline{e}^d\langle\boldsymbol{\xi}\rangle , \qquad (31)$$

where $\underline{e}^d\langle\boldsymbol{\xi}\rangle = \underline{e}\langle\boldsymbol{\xi}\rangle - \frac{\theta_\mathbf{x}\xi}{3}$, $\underline{e}\langle\boldsymbol{\xi}\rangle = \|\boldsymbol{\xi} + \boldsymbol{\eta}\| - \|\boldsymbol{\xi}\|$.

## 4. Cauchy stress recovery

For small strain deformation, the Cauchy stress tensor can be recovered by

$$\boldsymbol{\sigma}_\mathbf{x} = -\frac{m_\mathbf{x}}{3}\int_{H_\mathbf{x}} \mathbf{f_{x'x}} \otimes \boldsymbol{\xi}_r dV_{\mathbf{x}'} \cdot \mathbf{K}_\mathbf{x}^{-1} \qquad (32)$$

where $\mathbf{K}_\mathbf{x}$ is given by

$$\mathbf{K}_\mathbf{x} = \int_{H_\mathbf{x}} \omega(\boldsymbol{\xi})\,\boldsymbol{\xi}_r \otimes \boldsymbol{\xi}_r dV_{\mathbf{x}'} \qquad (33)$$

In fact, Eq.(32) is the same as the Hooke's law for isotropic elastic material.

$$\begin{aligned}
\boldsymbol{\sigma}_{\mathbf{x}} &= -\frac{m_{\mathbf{x}}}{3}\int_{H_{\mathbf{x}}} \mathbf{f}_{\mathbf{x}'\mathbf{x}} \otimes \boldsymbol{\xi}_r \mathrm{d}V_{\mathbf{x}'} \cdot \mathbf{K}_{\mathbf{x}}^{-1} \\
&= \frac{m_{\mathbf{x}}}{3}\int_{H_{\mathbf{x}}} \frac{3}{m_{\mathbf{x}}}\left((K-\frac{2\mu}{3})\theta_{\mathbf{x}}\omega(\boldsymbol{\xi})\boldsymbol{\xi}_r + 2\mu\omega(\boldsymbol{\xi})\boldsymbol{\eta}_r\right) \otimes \boldsymbol{\xi}_r \mathrm{d}V_{\mathbf{x}'} \cdot \mathbf{K}_{\mathbf{x}}^{-1} \\
&= \int_{H_{\mathbf{x}}} \left((K-\frac{2\mu}{3})\theta_{\mathbf{x}}\omega(\boldsymbol{\xi})\boldsymbol{\xi}_r + 2\mu\omega(\boldsymbol{\xi})\boldsymbol{\eta}_r\right) \otimes \boldsymbol{\xi}_r \mathrm{d}V_{\mathbf{x}'} \cdot \mathbf{K}_{\mathbf{x}}^{-1} \\
&= \int_{H_{\mathbf{x}}} \left((K-\frac{2\mu}{3})\theta_{\mathbf{x}}\omega(\boldsymbol{\xi})\boldsymbol{\xi}_r \otimes \boldsymbol{\xi}_r + 2\mu\omega(\boldsymbol{\xi})\boldsymbol{\eta}_r \otimes \boldsymbol{\xi}_r\right) \mathrm{d}V_{\mathbf{x}'} \cdot \mathbf{K}_{\mathbf{x}}^{-1} \\
&= \left((K-\frac{2\mu}{3})\theta_{\mathbf{x}}\int_{H_{\mathbf{x}}} \omega(\boldsymbol{\xi})\boldsymbol{\xi}_r \otimes \boldsymbol{\xi}_r \mathrm{d}V_{\mathbf{x}'} + 2\mu\int_{H_{\mathbf{x}}} \omega(\boldsymbol{\xi})\boldsymbol{\eta}_r \otimes \boldsymbol{\xi}_r \mathrm{d}V_{\mathbf{x}'}\right) \cdot \mathbf{K}_{\mathbf{x}}^{-1} \\
&= \left((K-\frac{2\mu}{3})\theta_{\mathbf{x}}\int_{H_{\mathbf{x}}} \omega(\boldsymbol{\xi})\boldsymbol{\xi}_r \otimes \boldsymbol{\xi}_r \mathrm{d}V_{\mathbf{x}'} + 2\mu\int_{H_{\mathbf{x}}} \omega(\boldsymbol{\xi})\boldsymbol{\varepsilon}_{\mathbf{x}} \cdot \boldsymbol{\xi}_r \otimes \boldsymbol{\xi}_r \mathrm{d}V_{\mathbf{x}'}\right) \cdot \mathbf{K}_{\mathbf{x}}^{-1} \\
&= \left((K-\frac{2\mu}{3})\theta_{\mathbf{x}}\int_{H_{\mathbf{x}}} \omega(\boldsymbol{\xi})\boldsymbol{\xi}_r \otimes \boldsymbol{\xi}_r \mathrm{d}V_{\mathbf{x}'} + 2\mu\boldsymbol{\varepsilon}_{\mathbf{x}} \cdot \int_{H_{\mathbf{x}}} \omega(\boldsymbol{\xi})\boldsymbol{\xi}_r \otimes \boldsymbol{\xi}_r \mathrm{d}V_{\mathbf{x}'}\right) \cdot \mathbf{K}_{\mathbf{x}}^{-1} \\
&= \left((K-\frac{2\mu}{3})\theta_{\mathbf{x}} \cdot \mathbf{K}_{\mathbf{x}} + 2\mu\boldsymbol{\varepsilon}_{\mathbf{x}} \cdot \mathbf{K}_{\mathbf{x}}\right) \cdot \mathbf{K}_{\mathbf{x}}^{-1} \\
&= (K-\frac{2\mu}{3})\theta_{\mathbf{x}} + 2\mu\boldsymbol{\varepsilon}_{\mathbf{x}} \\
&= K\theta_{\mathbf{x}} + 2\mu(\boldsymbol{\varepsilon}_{\mathbf{x}} - \frac{\theta_{\mathbf{x}}}{3}\mathbf{I}),
\end{aligned} \qquad (34)$$

where $\mathbf{I}$ is the identity matrix. It can be seen that under small strain deformation, the Cauchy stress tensor is recoverable.

## 5. Three type of fractures

There are general two types of fractures, brittle fracture and the ductile fracture, and much effort is exerted by the researchers [33, 34, 35, 36]. On the other hand, based on the crack separation modes, there are three types of cracks, i.e. mode I crack, mode II crack and mode III crack. As the ductile fracture is closely related to the complicated material constitutions, we only discuss the brittle fractures in three modes.

### 5.1. Critical deviatoric bond strain damage rule

In the implementation of the peridynamics, fracture is introduced by removing particles from the neighbor list once the bond stretch exceeds the critical value. In order to specify whether a bond is broken or not, a history-dependent scalar valued function $\mu$ is introduced [12],

$$\mu(t, \boldsymbol{\xi}) = \begin{cases} 1 & \text{if } s(t', \boldsymbol{\xi}) < s_0 \text{ for all } 0 \leq t' \leq t, \\ 0 & \text{otherwise.} \end{cases} \quad (35)$$

Then, based on function $\mu(t, \boldsymbol{\xi})$, the local damage at $\mathbf{x}$ is defined as

$$\phi(\mathbf{x}, t) = 1 - \frac{\int_{H_\mathbf{x}} \mu(\mathbf{x}, t, \boldsymbol{\xi}) \mathrm{d}V_\xi}{\int_{H_\mathbf{x}} \mathrm{d}V_\xi} \ . \quad (36)$$

The damage rule of Eq.(35) does not account for the shear deformation but only the extensional deformation of the bond, thus is not applicable in the current peridynamic formulation with shear deformation. The reason is that Mode I crack forms symmetric crack surfaces, while those by Mode II or Mode III are antisymmetric. For mode I crack, fracture surfaces $S^+$ and $S^-$ are symmetric with respected to the dash line as shown in Fig.4(a). The traditional peridynamics damage rule of critical extension works well in such situation. Given the displacement in Fig.4(a), both bond $ij$ and bond $il$ are under tension, the breakage of these bonds contributes to the formation of symmetric fracture surfaces.

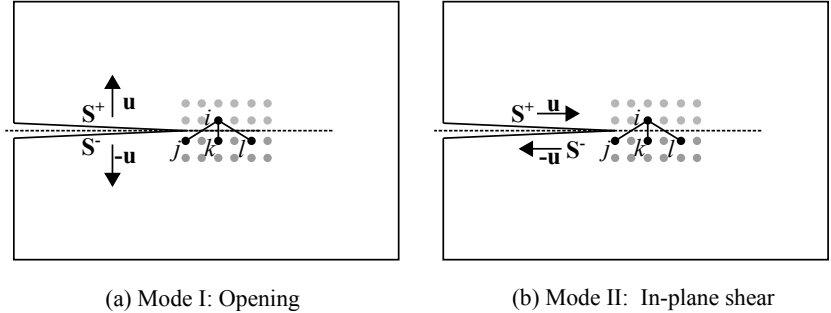

(a) Mode I: Opening    (b) Mode II: In-plane shear

Figure 4: Mode I and Mode II cracks in peridynamics.

However, in the mode II crack, the maximal extension rule is no longer valid. We note that for shear fracture, the deformation of fracture surfaces are anti-symmetric with respect to the dashed line as shown in Fig.4(b). Bond $ij$ is under tension while bond $il$ under compression thought they almost have the same shear bond strain. If the maximal extension rule is adopted, bond $ij$ is broken but bond $il$ is kept. Therefore, the maximal extension rule fails in mode II crack. The analysis also applies to the Mode III crack.

As the shear deformation accounts for the shear damage, the damage rule should be at least related to the shear deformation. Since bond $ik$ has the maximal shear deformation, it is reasonable to cut the bond $ik$ prior to bond $ij$ and bond $il$. It is necessary to design a new damage rule which is both compatible with the tensile damage and the shear damage. We notice some facts, for mode I cracks, particles in $S^+$ or in $S^-$ are under tension; for mode II and III cracks, particle in $S^+$ is under compression while particle in $S^-$ is under tension. Hence, we propose a maximal deviatoric bond strain damage rule as

$$\mu(t, \boldsymbol{\xi}) = \begin{cases} 0 & \text{if } \varepsilon^d(t', \boldsymbol{\xi}) \geq \varepsilon^d_{max} \text{ and } (\theta_{\mathbf{x}} \geq 0 \text{ or } \theta_{\mathbf{x}'} \geq 0), \text{ for all } 0 \leq t' \leq t, \\ 1 & \text{otherwise,} \end{cases} \quad (37)$$

where $\varepsilon^d = \|\boldsymbol{\varepsilon}^d\|$, $\theta_{\mathbf{x}}$ is the volume strain for particle $\mathbf{x}$. The damage happens when the the deviatoric bond strain exceeds critical value and at least one particle's volume strain is under tension.

5.2. Energy release rate to critical deviatoric bond strain

The volume strain is related to the hydrostatic pressure, and the anti-symmetric fracture surface in Mode II/III crack depends more on the deviatoric strain than the volume strain; therefore, we assume the volume strain along the crack surface is zero. Based on Eq. (22) and Eq.(26), the potential elastic energy for each bond in 3-D is

$$e^d_{max} = \frac{3\mu}{m}\omega(\boldsymbol{\xi})\boldsymbol{\eta}^d \cdot \boldsymbol{\eta}^d = \frac{3\mu}{m}(\varepsilon^d_{max})^2 \quad (38)$$

where $w(\boldsymbol{\xi}) = \xi^{-2}$, $m = \int_H w(\boldsymbol{\xi})\boldsymbol{\xi}_r \cdot \boldsymbol{\xi}_r \, dV = \frac{4\pi\delta^3}{3}$. Following the similar routine in [12], based on Fig.5, the work $G_0$ required to break all the bonds per unit fracture area is

$$\begin{aligned} G_0 &= \int_0^\delta \int_0^{2\pi} \int_z^\delta \int_0^{\cos^{-1} z/\xi} e^d_{max} \xi^2 \sin\phi \, d\phi \, d\xi \, d\theta \, dz \\ &= \frac{\pi\delta^4}{4}e^d = \frac{3\mu\pi\delta^4}{4m}(\varepsilon^d_{max})^2. \end{aligned} \quad (39)$$

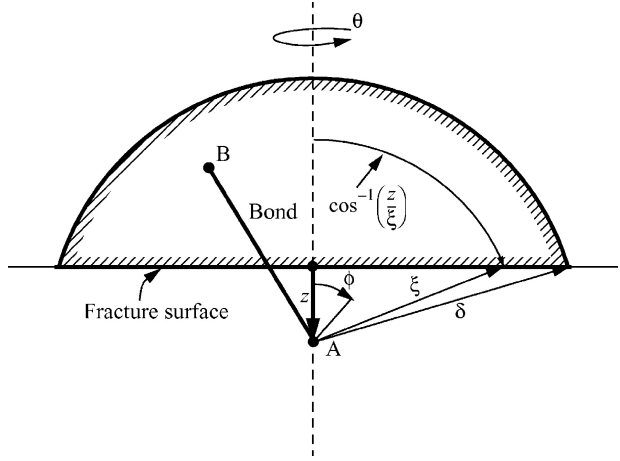

Figure 5: The energy release rate to create the crack surface [12].

Then, the critical bond strain $\varepsilon_{max}^d$ can be expressed as

$$\varepsilon_{max}^d = \frac{4}{3}\sqrt{\frac{G_0}{\mu\delta}} \qquad (40)$$

## 6. Numerical examples

### 6.1. Two-dimensional wave reflection in a rectangular plate

Consider a rectangular plate with dimensions of $0.1\times0.04$ m$^2$ (see Fig. 6). The Young's modulus, density and the Poisson's ratio used for the plate are $E = 1$, $\rho = 1$ and $\nu = 0$, respectively. Note that this is actually a 1-D problem solved in 2D. The initial displacement applied to the plate is described by the following equation

$$u_0(x,y) = 0.0002\exp[-(\frac{x}{0.01})^2], \quad v_0(x,y) = 0, \quad x \in [0, 0.1], y \in [0, 0.04], \qquad (41)$$

where $u_0$ and $v_0$ denote the displacement in the $x$ and $y$ directions respectively. The wave speed is $v = \sqrt{E/\rho} = 1$ m/s. The plate is discretized with a particle size of $\Delta x = 1e-3$m excerpt the right top part of the plate which is discretized with $\Delta x = 5e-4$ m. Each particle's horizon radius is set as $\delta_{\mathbf{x}} = 3.015\Delta x$. There is a sharp transition at $x = 0.05$ m.

Fig.7(a) shows the displacement wave was approaching the interface of different horizon size at $t = 0.0315s$. Some variance is observed at location $x = 0$, which is due to the non-local boundary effect of peridynamics. When the wave passed the interface at $t = 0.0735s$, there is no obvious reflection wave, as shown in Fig.7(b). Hence, the sharp transition of horizon size has little influence on the current dual-horizon peridynamics. At $t = 0.1s$, the wave reached the place $x = 0.1m$ and was reflected back at this free boundary. At $t = 0.01365 \sim 0.01575s$, the reflected wave passed through the interface of different horizon sizes, and no obvious reflected wave is observed, as shown in Fig.7(c) and Fig.7(d).

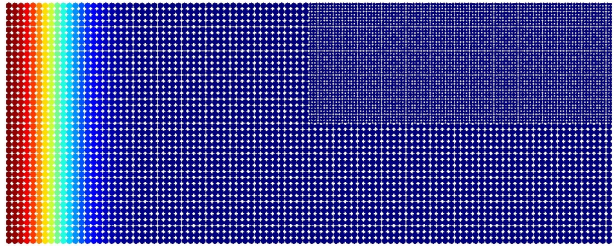

Figure 6: The particle distribution of the plate.

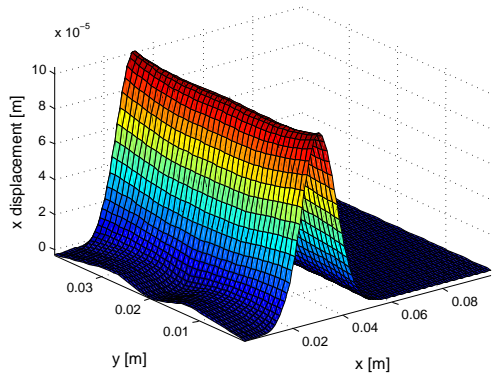

(a) $t = 0.0315s$

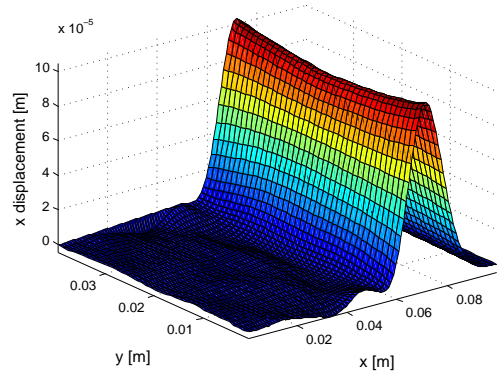

(b) $t = 0.0735s$

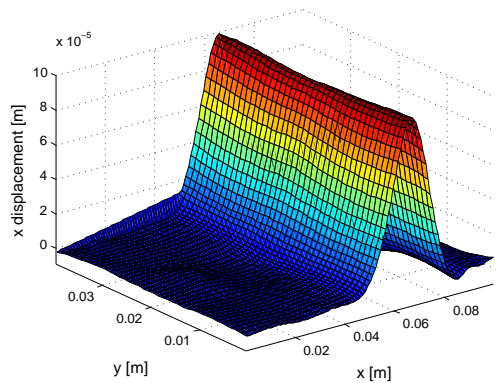

(c) $t = 0.1365s$

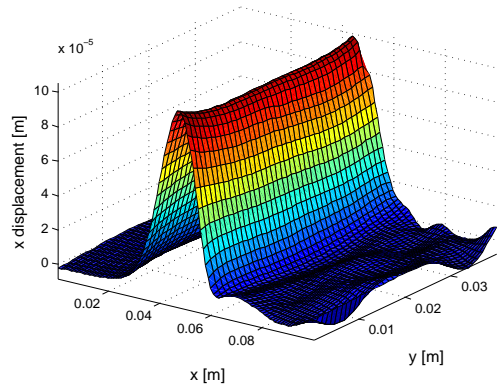

(d) $t = 0.1575s$

Figure 7: The profiles of $x$ displacement at different time.

*6.2. Large rigid rotation test*

The rigid rotation should result no internal stress. We test the large rigid rotation in 2D plate (Case I) and 3D plate (Case II). In Case I, we rotate a 2D plate of 1m×1m with an angular velocity of 15° per step and calculate the internal stress with respect to the initial configuration. In Case II, a 3D plate of 1m×1m×0.2m, is rotated along the Euler axis of (1,1,1) with an angular velocity of 30° per step. The parameters for material for both plates are $E = 30$GPa, $\nu = 1/3$, $\rho = 2400$ kg/$m^3$. The stress recovery method is employed to calculate the stress. Here are some results for stress component $\sigma_{11}$. The maximal value of $\sigma_{11}$ in 2D plate is within 3e-3 Pa, and that in 3D plate is within 10 Pa, both which are due to the numerical error in calculating the rotation matrix and therefore are neglectable compared with the large rigid rotations. It is safe to conclude that the rigid rotation is subtracted from the total deformation in the framework of current peridynamic formulation.

*6.3. 2D plate with hole*

Consider a 2D plate of 1m×1m with a hole of $r = 0.125$ m and force boundary condition of $P = 1$ MPa, as shown in Fig.9, the problem is solved by the current peridynamics and Abaqus standard. Several simulations on different mesh on Abaqus are compared, and we choose the converged results. The plate is discretized into 38,024 particles with particle spacing of $\Delta x = 0.005$m. The parameters for material are $E = 30$GPa, $\nu = 1/3$, $\rho = 2400$ kg/$m^3$. In order to obtain the static solution, a damping coefficient $\alpha = 1e4$ is used, where the damping force for particle $\mathbf{x}$ is calculated by

$$\mathbf{f}_{\mathbf{x}}^{damp} = -\alpha \mathbf{v}_{\mathbf{x}} m_{\mathbf{x}}, \tag{42}$$

where $m_{\mathbf{x}}$ is the particle mass.

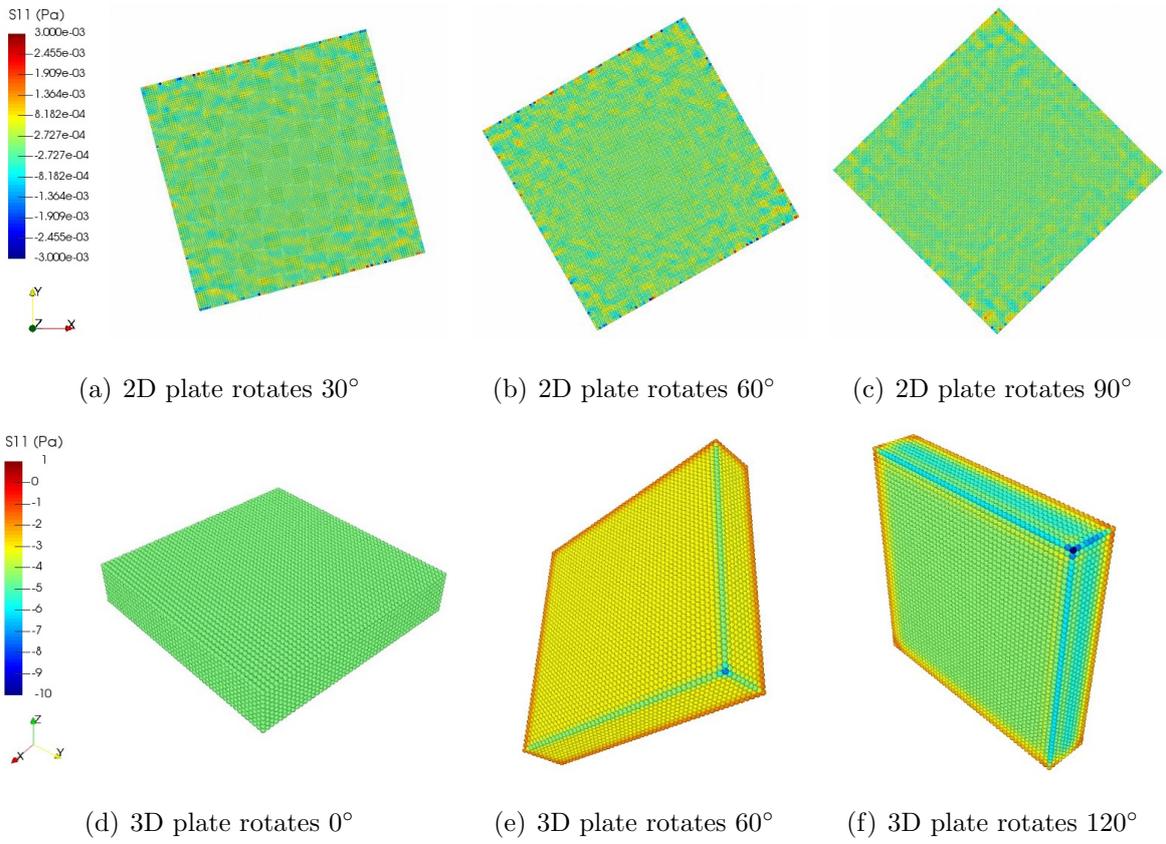

(a) 2D plate rotates 30°  (b) 2D plate rotates 60°  (c) 2D plate rotates 90°

(d) 3D plate rotates 0°  (e) 3D plate rotates 60°  (f) 3D plate rotates 120°

Figure 8: The internal stress $\sigma_{11}$ on different rotation angles.

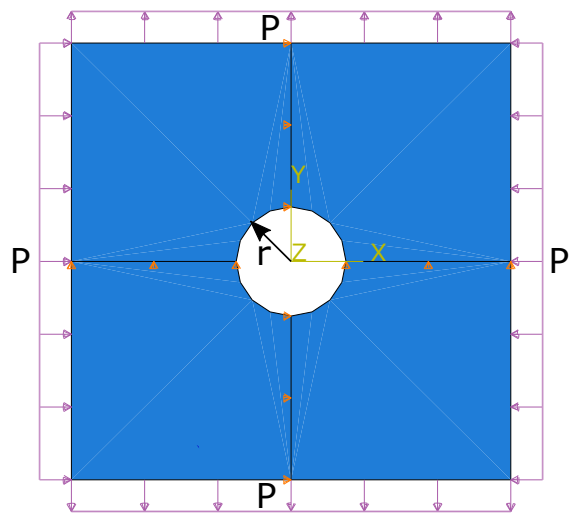

Figure 9: The setup of the plate.

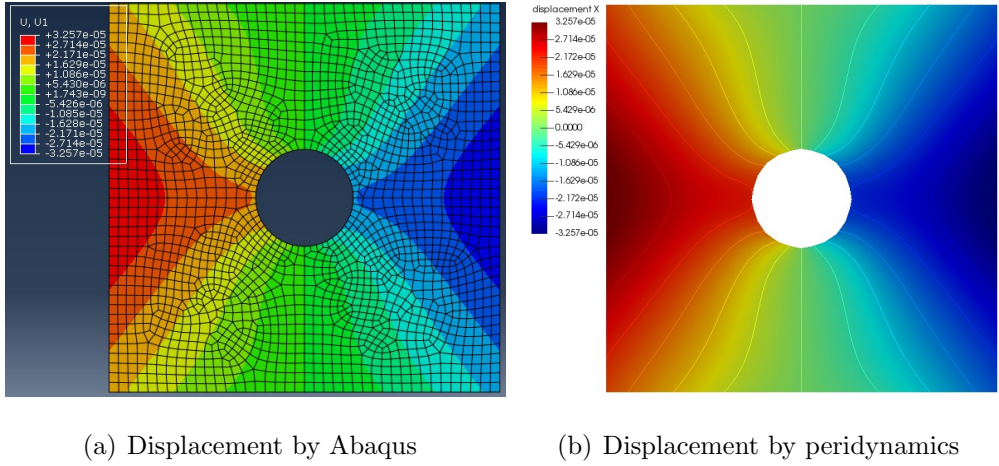

(a) Displacement by Abaqus  (b) Displacement by peridynamics

Figure 10: The displacement in $x$ direction by peridynamics and Abaqus.

Fig.10(a,b) shows that the displacement field by peridynamics agrees well with FEM results since the contours are plotted with almost the same tick labels.

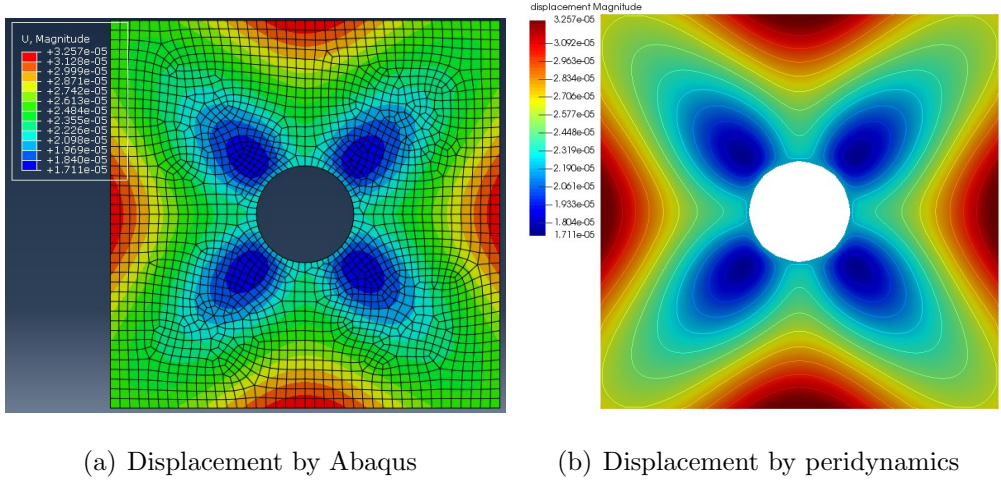

(a) Displacement by Abaqus  (b) Displacement by peridynamics

Figure 11: The displacement field by peridynamics and Abaqus.

The comparison in Fig.13 shows that the current peridynamics agrees well with the results by Abaqus Standard.

As shown in Table. 1, and take the results by Abaqus standard as the reference, the maximal error for displacement is 2.95%, for $\sigma_{xx}$ 28.3% and for $\sigma_{yy}$ 29.7%, which all happened on the boundaries. The reason is the non-local effect of peridynamics. For particle far away from the boundaries, the good agreement is observed as shown in Figs.10,11,12 and 13.

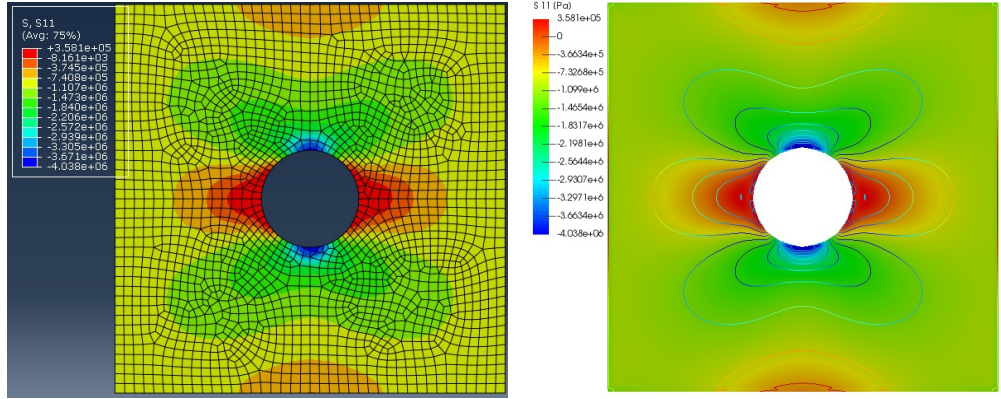

(a) $\sigma_{xx}$ by Abaqus
(b) $\sigma_{xx}$ by peridynamics

Figure 12: The stress $\sigma_{xx}$ by peridynamics and Abaqus.

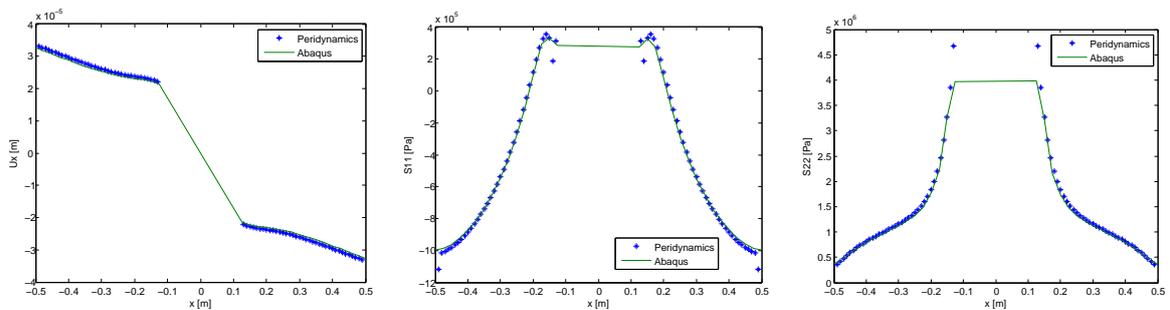

(a) $u_x$
(b) $\sigma_{xx}$
(c) $\sigma_{yy}$

Figure 13: The comparison of $u_x$, $\sigma_{xx}$ and $\sigma_{yy}$ on section $y = 0$. The relative errors are $\text{RE}(u_x) = 1.63\%$ on the boundary, $\text{RE}(\sigma_{xx}) = 12.05\%$ on the boundary, $\text{RE}(\sigma_{yy}) = 17.14\%$ on the hole, where $\text{RE}(v) = |\frac{v_{\text{peridynamics}}}{v_{\text{Abaqus}}} - 1|$.

| Model | $u_{xmax}$ | $u_{max}$ | $\|\sigma_{xx}\|_{max}$ | $\|\sigma_{yy}\|_{max}$ |
|---|---|---|---|---|
| peridynamics | 3.35e-5 m | 3.35e-5 m | 5.18 MPa | 5.18 Pa |
| Abaqus | 3.26e-5 m | 3.26e-5 m | 4.04 MPa | 3.99 MPa |
| peridynamics/Abaqus−1 | 2.95e-2 | 2.95e-2 | 0.283 | 0.297 |

Table 1: The comparison of peridynamics and Abaqus.

*6.4. Mode I,II,III cracks*

As the current peridynamics concentrates on the linear elastic material, we assume the fractures are brittle. We simulate three types of brittle fractures with the damage rule given in Eq.(40). The material parameters are $E = 30$GPa, $\nu = 1/3$, $\rho = 2400$ kg/$m^3$, $G_0 = 200 J/m^2$. The dimensions are $0.5 \times 0.2 \times 0.1 m^3$ with initial crack of length 0.3 m, and three types of velocity boundary conditions corresponding to three modes of fractures are applied, respectively, as shown in Fig.14. The velocity boundary of $v = 1m/s$ is applied on one layer of particles on the surface. The specimen is discretized into 80,000 particles with particle spacing of 5e-3 m. The time increment is 8.09e-7 s. The averaged crack propagation speeds for three modes of fracture, respectively, are 343.6m/s, 824.4 m/s and 265.9 m/s, which are all smaller than the Rayleigh wave speed. The snapshots of different mode crack are shown in Figs.15,16,17. The crack propagation speed is shown in Figs.18,19,20.

For mode I crack, the crack starts to propagate at $t = 385\mu s$, and a peak is observed; with the release of strain energy, the crack speed decreases to zero at $t = 590\mu s$. As the deformation going on, the crack restarts to propagate at $t = 630\mu s$ and the crack velocity increases gradually.

For mode II crack, the displacement in $x$ direction contributes to the bond strain in $x$ direction. The strain exceeds the critical bond strain quickly, so the crack starts to propagate at $t = 120\mu s$. In a very short time, the specimen is separated into two pieces.

For mode III crack, under the condition of the out-of-plane displacement field, the stress concentration firstly happens on the up and down of the crack tip. The crack starts to propagate at $t = 360\mu s$. A crack front edge in shape of concave arc is observed during the crack propagation. The crack speed shows some kind of periodicity. The formation of crack surface consumes the strain energy, and the accumulated strain energy stimulates the crack propagation.

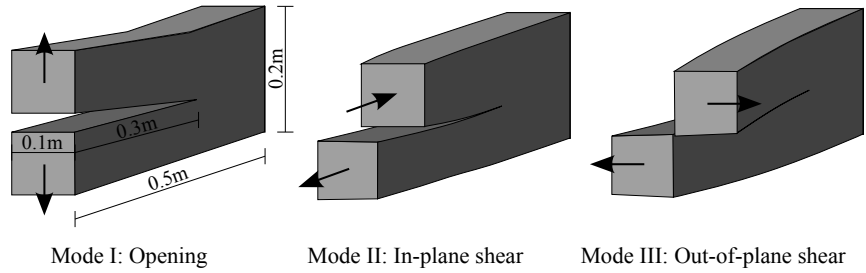

Figure 14: The setup of the specimens and the velocity boundary conditions.

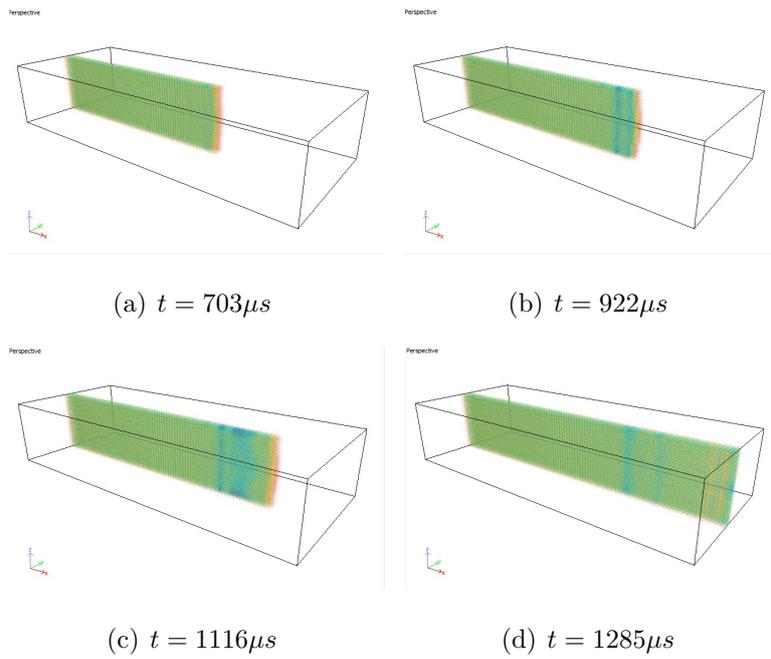

(a) $t = 703\mu s$

(b) $t = 922\mu s$

(c) $t = 1116\mu s$

(d) $t = 1285\mu s$

Figure 15: The propagation of Mode I crack.

## 7. Conclusion

In the paper, we have proposed a new peridynamic formulation with shear bond deformation. The key step is to deduce the rigid rotation of the total deformation. Compared with BB peridynamics and OSB peridynamics, current peridynamics is more similar to the classic elastic mechanics and allows the stress to be recovered while there is no stress in BB peridynamics and OSB peridynamics. On the other hand, compared with the NOSB peridynamics, the current peridynamics is simpler in expression. The current formulation allows for arbitrary rotations.

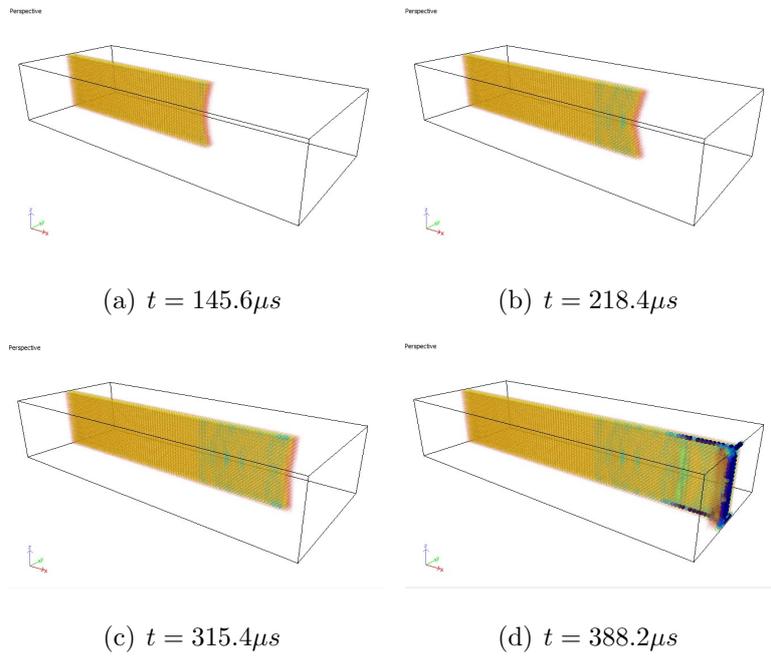

(a) $t = 145.6\mu s$
(b) $t = 218.4\mu s$
(c) $t = 315.4\mu s$
(d) $t = 388.2\mu s$

Figure 16: The propagation of Mode II crack.

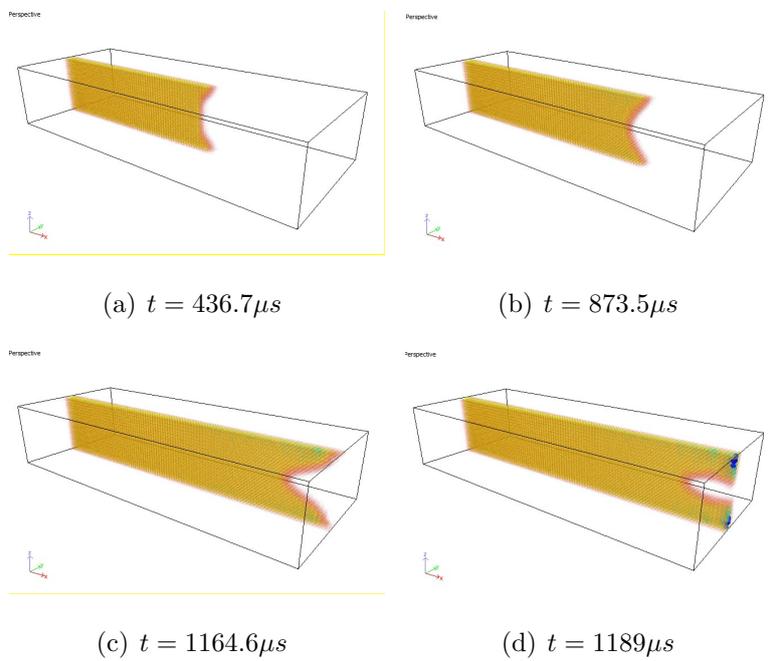

(a) $t = 436.7\mu s$
(b) $t = 873.5\mu s$
(c) $t = 1164.6\mu s$
(d) $t = 1189\mu s$

Figure 17: The propagation of Mode III crack.

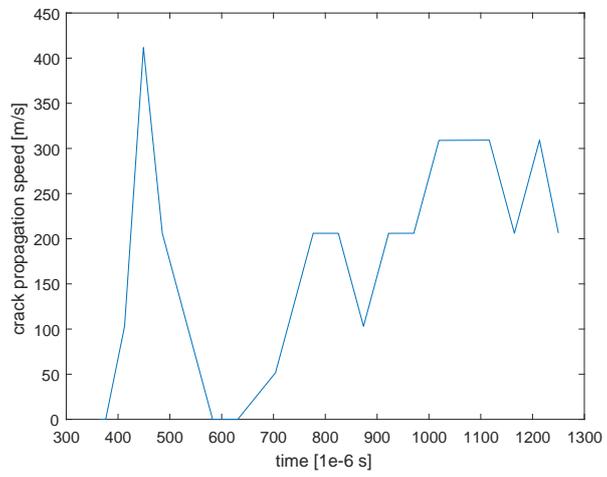

Figure 18: The crack propagation speed of Mode I crack.

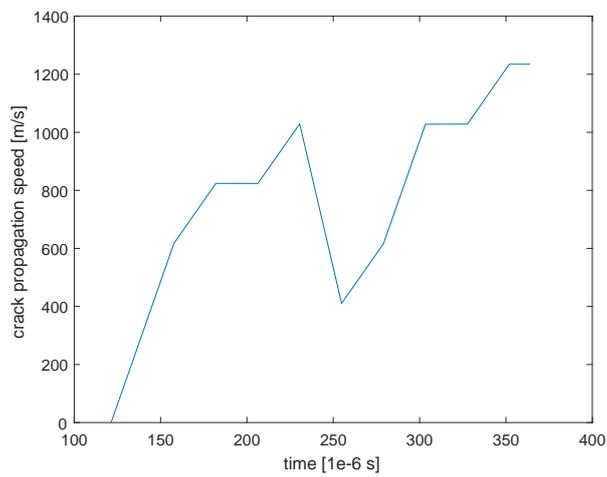

Figure 19: The crack propagation speed of Mode II crack.

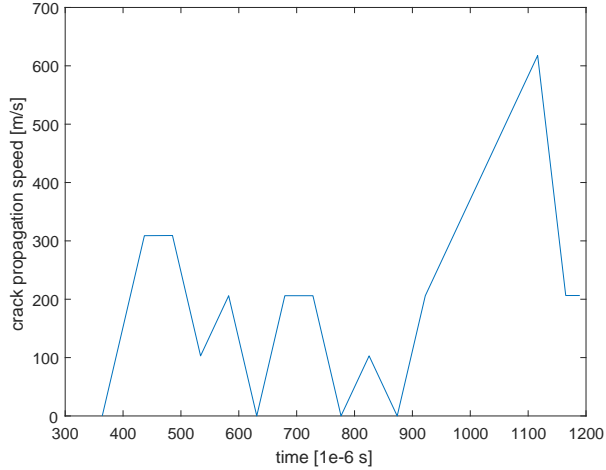

Figure 20: The crack propagation speed of Mode III crack.

We also have proposed a new damage rule based on the maximal deviatoric bond strain. This damage rule is compatible with three modes of brittle fractures. Compared with the traditional maximal bond stretch, the current damage rule can predict the in-plane shear and out-of-plane shear fracture with ease. Such damage rule is possible to be applied in NOSB peridynamics for further research.

The accuracy in solving continuous problem and capability in fracture modeling of current peridynamic formulation have been demonstrated by several examples.

## Acknowledgements

The authors acknowledge the supports from FP7 Marie Curie Actions ITN-INSIST and IIF-HYDROFRAC (623667), the National Basic Research Program of China (973 Program: 2011CB013800) and NSFC (51474157), the Ministry of Science and Technology of China (Grant No.SLDRCE14-B-28, SLDRCE14-B-31).

cording to von mises yield criteria with isotropic hardening. *Journal of the Mechanics and Physics of Solids*, 86:192 – 219, 2016.

[17] J Amani, E Oterkus, P Areias, Goangseup Zi, T Nguyen-Thoi, and T Rabczuk. A non-ordinary state-based peridynamics formulation for thermoplastic fracture. *International Journal of Impact Engineering*, 87:83–94, 2016.

[18] Abigail Agwai, Ibrahim Guven, and Erdogan Madenci. Fully coupled peridynamic thermomechanics. *53rd AIAA Structures, Structural Dynamics and MaterialsConference, Honolulu, Hawaii*, 2012.

[19] Florin Bobaru and Monchai Duangpanya. A peridynamic formulation for transient heat conduction in bodies with evolving discontinuities. *Journal of Computational Physics*, 231(7):2764–2785, 2012.

[20] Florin Bobaru and Monchai Duangpanya. The peridynamic formulation for transient heat conduction. *International Journal of Heat and Mass Transfer*, 53(19):4047–4059, 2010.

[21] Shubhankar Roy Chowdhury, Pranesh Roy, Debasish Roy, and JN Reddy. A peridynamic theory for linear elastic shells. *arXiv preprint arXiv:1508.00082*, 2015.

[22] C. Diyaroglu, E. Oterkus, S. Oterkus, and E. Madenci. Peridynamics for bending of beams and plates with transverse shear deformation. *International Journal of Solids and Structures*, 6970:152 – 168, 2015.

[23] James OGrady and John Foster. Peridynamic plates and flat shells: a non-ordinary, state-based model. *International Journal of Solids and Structures*, 51(25):4572–4579, 2014.

[24] E. Oterkus, E. Madenci, O. Weckner, S. Silling, P. Bogert, and A. Tessler. Combined finite element and peridynamic analyses for predicting failure in a stiffened composite curved panel with a central slot. *Composite Structures*, 94:839–850, 2012.

[25] Qiang Du, Max Gunzburger, Richard B Lehoucq, and Kun Zhou. A nonlocal vector calculus, nonlocal volume-constrained problems, and nonlocal balance laws. *Mathematical Models and Methods in Applied Sciences*, 23(03):493–540, 2013.

[26] Ted Belytschko, Nicolas Moës, Shuji Usui, and Chandu Parimi. Arbitrary discontinuities in finite elements. *International Journal for Numerical Methods in Engineering*, 50(4):993–1013, 2001.

[27] Timon Rabczuk, Stphane Bordas, and Goangseup Zi. On three-dimensional modelling of crack growth using partition of unity methods. *Computers & Structures*, 88(2324):1391 – 1411, 2010. Special Issue: Association of Computational Mechanics  United Kingdom.

[28] I. Babuška and J.M. Melenk. The partition of the unity finite element method. Technical Report BN-1185, Institute for Physical Science and Technology, University of Maryland, Maryland, 1995.

[29] Huilong Ren, Xiaoying Zhuang, Yongchang Cai, and Timon Rabczuk. Dual-horizon peridynamics. *arXiv preprint arXiv:1506.05146*, 2015.

[30] Stewart Silling, David Littlewood, and Pablo Seleson. Variable horizon in a peridynamic medium.